\documentclass[reprint,superscriptaddress,amsmath,amssymb,aps,pra]{revtex4-2}

\usepackage{graphicx}
\usepackage{dcolumn}
\usepackage{bm}
\usepackage{epstopdf}
\usepackage{braket}
\usepackage{color}
\usepackage{mathrsfs}
\usepackage{ragged2e}
\usepackage{eqnarray}
\usepackage{subeqnarray}
\usepackage{cases}
\usepackage{nicefrac}
\usepackage{tikz}
\usepackage{float}
\usepackage[colorlinks=true,citecolor=blue,urlcolor=blue]{hyperref}
\usepackage{amsmath,amsfonts,amssymb,natbib}
\usepackage[standard]{ntheorem}
\usepackage{txfonts} 
\usepackage[all]{hypcap}

\begin{document}

\preprint{APS/123-QED}

\title{Universal atomic spin response to arbitrary periodic optical pumping}

\author{Ling-Yan Hu}
\thanks{These authors contributed equally to this work.}
\affiliation{School of Physics, Hefei University of Technology, Hefei, Anhui 230009, China.}%

\author{Ying-Hao Ye}
\thanks{These authors contributed equally to this work.}
\email{yyhustc@mail.ustc.edu.cn}
\affiliation{School of Physics, Hefei University of Technology, Hefei, Anhui 230009, China.}%

\author{Xin-Ru Liu}
\affiliation{School of Physics, Hefei University of Technology, Hefei, Anhui 230009, China.}%

\author{Dui-Gao Yi}
\affiliation{School of Physics, Hefei University of Technology, Hefei, Anhui 230009, China.}%

\author{Zhi-Fei Yu}
\email{zfyu-phy@hfut.edu.cn}
\affiliation{School of Physics, Hefei University of Technology, Hefei, Anhui 230009, China.}%
\affiliation{Key Laboratory of Opto-Electronic Information Acquisition and Manipulation (Anhui University), Ministry of Education, Hefei, Anhui 230039, China.}%

\author{Ting Zhang}
\affiliation{School of Physics, Hefei University of Technology, Hefei, Anhui 230009, China.}%

\author{Bing Chen}
\email{bingchenphysics@hfut.edu.cn}
\affiliation{School of Physics, Hefei University of Technology, Hefei, Anhui 230009, China.}


\date{\today}

\begin{abstract}
We investigate the atomic response to arbitrary periodic pumping in a Fourier--Floquet framework. The comparison between weak- and strong-pump conditions reveals a pronounced inter-harmonic Floquet mixing in the latter. Notably, anomalous linewidth narrowing is also observed, which is unlikely to originate from diffusion-induced Ramsey effects given the strong suppression of atomic diffusion by 760 Torr of $N_2$. A possible explanation for this narrowing lies in the probe-weighted superposition of complex responses from regions with different pump-induced relaxation rates, motivating an extension of the uniform distribution model. The resulting spatially averaged description provides substantially better agreement with the strong-pump measurements than either the bare Lorentzian or conventional Floquet model, as demonstrated by a controlled test using a tailored pump waveform, while all three models converge in the weak-pump limit. These results establish a broadly applicable framework for predicting atomic responses across different pump waveforms and pumping strengths, with potential for waveform-engineered sensing.
\end{abstract}

\maketitle

\section{\label{sec:introducetion}Introduction}
Coherent atomic spin dynamics arise from the interplay of Larmor precession, relaxation, and external driving, with temporal modulation of the excitation providing an additional means of control \cite{Fabricant2023review}. In optically pumped ensembles, periodically modulated light can synchronously drive spin precession and produce magnetic resonances at nonzero bias fields, as established by Bell and Bloom \cite{BellBloom57,BellBloom61}. Particularly, in suitable polarization arrangements, modulation near the Larmor frequency or twice that frequency can resonantly excite atomic orientation or alignment, respectively \cite{Lightellipticityalignment}. Such techniques are widely used in magnetometry \cite{np07OPMreview,Liu26review} and nonlinear magneto-optical spectroscopy \cite{Budker2002review}, with amplitude-, frequency-, and polarization-modulated excitation providing versatile means of probing and controlling atomic spin coherence \cite{AMGawlik06,FMBudker,PMOE2013,GrujicWeis2013}.

For a periodic pump waveform, Fourier decomposition provides a natural representation of the optical excitation \cite{Alex05review,GrujicWeis2013}. Higher Fourier harmonics can resonantly drive spin precession when their frequencies match the Larmor frequency, producing a sequence of subharmonic resonances \cite{PMOE2013,Breschi2013PRA}. In the weak-pump limit, the relative resonance strengths are directly related to the Fourier coefficients of the modulation \cite{GrujicWeis2013}, suggesting that individual pump harmonics act as approximately independent driving channels. However, the validity of this independent-harmonic picture at higher pump powers remains less clear \cite{Floquet2021PRA}.

Beyond the weak-pump limit, Floquet theory provides a natural framework for periodic spin dynamics \cite{Floquet65PR}. Such formulations have been developed for optically pumped systems with general periodic pumping profiles and can be extended to multiple radio-frequency fields \cite{Floquet2021PRA}. Related studies have examined nonquasistatic responses to time-dependent magnetic fields \cite{Bevilacqua2021}, as well as Floquet-mediated dynamics under stroboscopic microwave and bichromatic radio-frequency driving \cite{Florez24PRA,Geng26PRA}. Nevertheless, to the best of our knowledge, the connection between the imposed pump spectrum and the measured complex response of individual harmonic orders has not been systematically studied across the independent- and coupled-harmonic regimes. Moreover, spatially inhomogeneous optical pumping and detection cause the measured signal to combine differently weighted local responses \cite{Lightdistribution}. These considerations motivate a unified treatment of the pump spectrum, inter-harmonic Floquet mixing, and spatial averaging.

\begin{figure*}[t!]
	\centering
	\includegraphics[width=\linewidth]{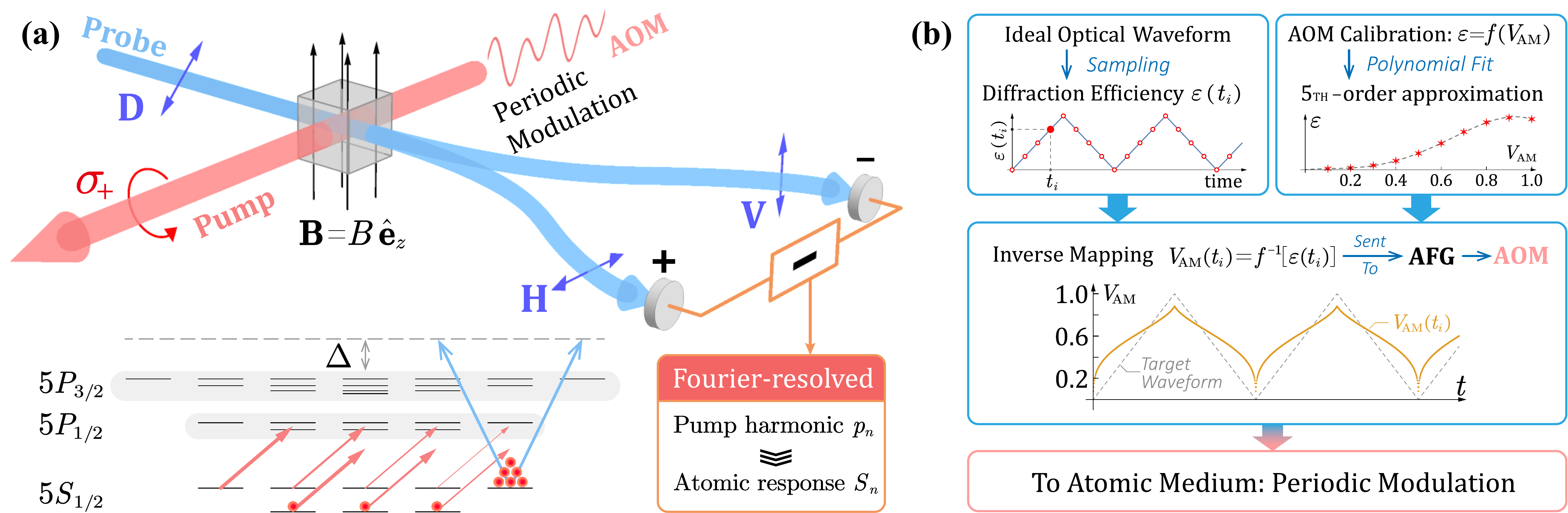}
	\caption{\textbf{Experimental principle.} 
		(a) Mutually orthogonal pump--probe--magnetic field geometry and relevant optical transitions. The periodically modulated circularly polarized pump drives the D1 transition, while the orthogonal linearly polarized probe, blue detuned by $\Delta$ from the D2 transition, detects the spin precession through polarization rotation.
		The pump harmonics and atomic response are then resolved in the Fourier domain. (b) Predistortion procedure for AOM modulation. The calibrated polynomial transfer function is inverted to generate the AFG script required for a prescribed modulation waveform.
	}
	\label{Fig:principle}
\end{figure*}
Harmonic-resolved phase measurements provide information complementary to resonance amplitudes. Comparing the full complex response across harmonic detuning therefore provides a more comprehensive assessment of the atomic dynamics and places stronger constraints on theoretical descriptions. Here, we investigate the harmonic-resolved complex atomic response to arbitrary periodic optical pumping. We develop a Fourier--Floquet formulation relating the measured pump spectrum to the atomic response, establishing the validity of the independent-harmonic approximation and incorporating spatially varying pumping and probe weighting. Measurements using six conventional waveforms characterize the weak- and strong-pump responses, while a tailored waveform with prescribed Fourier amplitudes and phases provides a controlled comparison of the bare Lorentzian, spatially uniform Floquet, and spatially averaged Floquet descriptions. Using measured pump spectra and independently determined relaxation rates as inputs, the spatially averaged model provides substantially better agreement with the strong-pump measurements. The resulting framework offers a predictive basis for waveform engineering and harmonic-resolved sensing, with potential applications to magnetic \cite{np07OPMreview}, radio-frequency \cite{Xiao24RadioMeasurement}, and microwave field measurements \cite{MicrowaveFieldMeasurementPRL,PRAppliedMicrowaveMeasurement}.

This article is organized as follows. Sec.~\ref{sec:principle} presents the scheme for periodic pumping and the characterization of modulated waveforms with distinct Fourier spectra. Sec.~\ref{sec:FourierFloquet} develops the Fourier--Floquet framework. Sec.~\ref{sec:sixwaveform} experimentally compares the independent-harmonic and Floquet-mixing regimes under weak and strong pumping conditions, revealing a linewidth behavior that motivates the spatial treatment introduced in Sec.~\ref{sec:spatialparameter}. Finally, Sec.~\ref{sec:engineeredwaveform} employs a deliberately engineered waveform to compare the different theoretical descriptions and examine the interplay between Floquet mixing and spatial averaging.

\section{\label{sec:principle}Experimental Scheme}
The experimental principle is depicted in Fig.~\ref{Fig:principle}. We employ a mutually orthogonal pump--probe--field geometry, as shown in Fig.~\ref{Fig:principle}(a). A circularly polarized pump beam propagating along the $x$ axis drives the D1 transition, while a static magnetic field $B_0\hat{e}_z$ defines the Larmor-precession axis. The measurements are performed in an uncoated $^{87}$Rb vapor cell containing 760 Torr of $N_2$ buffer gas. The high buffer-gas pressure strongly suppresses atomic diffusion across the optical beams, thereby suppressing transport-induced effects \cite{Pouliot2021PRA}, including Ramsey narrowing \cite{RamseyNarrowing06}.

The probe propagates along $\hat{e}_y$ and is blue detuned by approximately $\Delta=30~\mathrm{GHz}$ from the D2 transition. Before entering the vapor cell, it is linearly polarized along the diagonal ($D$) direction. Atomic spin polarization produces circular birefringence for the $\sigma_{\pm}$ components of the probe field, giving rise to a polarization rotation $\theta=kL(n_+-n_-)/2$. After passing through the cell, the probe is projected onto the horizontal ($H$) and vertical ($V$) polarization bases and detected differentially. For small rotation angles, $I_H-I_V\simeq2I_0\theta\propto P_y$, providing a direct readout of the spin-polarization component along the probe direction \cite{Fabricant2023}.

Another key element of the scheme is the controlled modulation of pump waveforms, realized by the Amplitude-modulation (AM) port of an AOM, as illustrated in Fig.~\ref{Fig:principle}(b). To compensate for the nonlinear dependence of the AOM diffraction efficiency $\varepsilon$ on the control voltage $V_{\rm AM}$, we employed a polynomial predistortion procedure. The calibrated transfer relation $\varepsilon=f(V_{\rm AM})$ was fitted with a fifth-order polynomial, and the required drive waveform was obtained from $V_{\rm AM}(t_i)=f^{-1}\!\left[\varepsilon(t_i)\right]$ within the calibrated operating range. The calculated voltage waveform was then replayed by the arbitrary function generator (AFG) to drive the AOM and the diffracted pump light was monitored by a fast-photondetector.
\begin{figure*}[ht!]
	\centering
	\includegraphics[width=\linewidth]{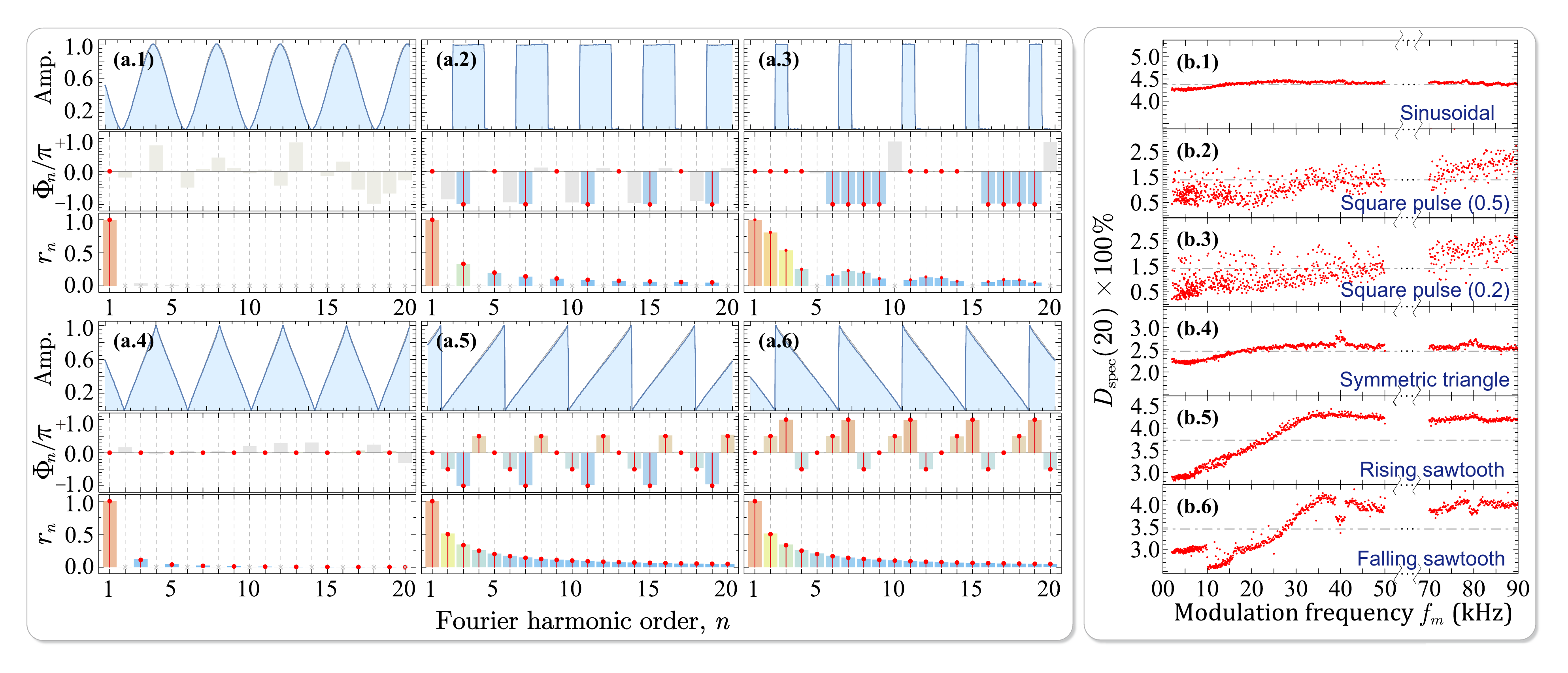}
	\caption{\textbf{Characterization of the generated periodic pump waveforms.} (a.1)--(a.6) Corresponding Fourier spectra
		for six representative periodic waveforms with distinct harmonic
		amplitude and phase structures at $f_m=40~\mathrm{kHz}$. The light-blue shaded profiles show the measured pump waveforms, while
		the gray ones denote the corresponding targets.
		The lower panels compare the measured relative harmonic amplitudes
		$r_n$ and phases $\Phi_n$ with the theoretical values (red markers). (b.1)--(b.6) Spectral deviation $D_{\rm spec}(20)$ as a function of
		modulation frequency for the corresponding waveforms, characterizing
		the operating bandwidth of the predistortion.}
	\label{Fig:differentwaveform}
\end{figure*}

To quantify the generated waveform, each registered trace was phase-folded
and averaged over multiple modulation cycles. Its complex Fourier coefficients
were calculated as $C_n=M^{-1}\sum_{k=0}^{M-1}v_k\exp(-i2\pi nk/M)$, with $M$ uniformly sampled points per period. The DC component $C_0$ corresponds to the cycle-averaged pump signal. The amplitude and phase of the $n$th harmonic ($n\geq1$) are then given by $A_n=2\lvert C_n \rvert$ and $\phi_n=\arg(C_n)$, respectively. We then define the relative amplitude and phase as
\begin{equation}
	r_n = A_n / A_1,
	\qquad
	\Phi_n=\arg\left\{
	\exp\left[i(\phi_n-n\phi_1) \in(-\pi,\pi]\right]
	\right\},
\end{equation}
to characterize the time origin-independent harmonic properties. To this end, the overall fidelity is evaluated by the normalized complex spectral distance between the experimental and ideal waveform:
\begin{equation}
	\label{eqn:Dspec}
	D_{\mathrm{spec}}(N)=
	\left[
	{
		\displaystyle
		\sum\nolimits_{n=1}^{N}
		\left|
		q_n^{\mathrm{exp}}-q_n^{\mathrm{th}}
		\right|^2
	}\Bigg/{
		\displaystyle
		\sum_{n=1}^{N}
		\left|
		q_n^{\mathrm{th}}
		\right|^2
	}
	\right]^{1/2}.
\end{equation}
where $N$ is the truncation order and $q_n=r_n e^{i\Phi_n}$ is the normalized complex harmonic coefficient, thus $D_{\mathrm{spec}}$ quantifies deviations in both amplitude and relative phase.

As representative examples, we employed six periodic drives with distinct harmonic structures: sinusoidal, square pulses with duty cycles of 0.5 and 0.2, symmetric triangular, and rising- and falling-sawtooth waveforms. Figs.~\ref{Fig:differentwaveform}(a.1)--(a.6) show the measured waveforms and their corresponding Fourier spectra at $f_m=40~\mathrm{kHz}$, whereas Figs.~\ref{Fig:differentwaveform}(b.1)--(b.6) show the spectral deviations $D_{\rm spec}(20)$ as functions of $f_m$.

The measured Fourier spectra closely reproduce the characteristic features of the target waveforms. The sinusoidal modulation is dominated by the fundamental component, while the 50\%-duty-cycle square pulse and symmetric-triangular waveform exhibit the expected odd-harmonic spectra. The harmonic amplitudes decay as $1/n^2$ for the symmetric triangle and as $1/n$ for the sawtooth waveforms. The rising- and falling-sawtooth show nearly identical amplitude spectra but opposite relative-phase winding. For the 20\%-duty-cycle square pulse, the measured harmonic amplitudes follow the theoretical envelope. Complementary to $D_{\rm spec}$, the ratio $A_1/C_0$ quantifies the relative weight of the fundamental component with respect to the DC component. For all six waveforms, its relative deviation from the theoretical value does not exceed $5.0\%$, while $D_{\rm spec}(20)<4.4\times10^{-2}$, demonstrating high-fidelity reproduction of the complex harmonic structure. 

The residual deviations are primarily associated with the finite modulation bandwidth of the waveform-generation setup. As shown in Figs.~\ref{Fig:differentwaveform}(b), $D_{\rm spec}(20)$ generally increases with $f_m$, with a pronounced waveform dependence. For instance, the sinusoidal and symmetric-triangular drives exhibit only weak variations in $D_{\rm spec}$, whereas the square-pulse and, more prominently, the sawtooth waveforms show increasing spectral distortion at higher $f_m$. This trend is consistent with the stronger limitation imposed by the finite modulation bandwidth on waveforms containing appreciable higher-order harmonics. To represent the periodic drive experienced by the atoms accurately, the actual waveforms, rather than the ideal ones, are used in the subsequent analysis.

\section{\label{sec:FourierFloquet}Fourier–Floquet framework}
The periodic pump waveform can be expressed in terms of complex Fourier components as
\begin{equation}
	\label{eqn:FourierExpand}
	p(t)=\sum\nolimits_{n=-\infty}^{+\infty}p_n e^{in\omega_m t},
\end{equation}
where $\omega_m = 2\pi f_m$, $p_n=(\omega_m/2\pi)\int_0^{2\pi/\omega_m} p(t)e^{-in\omega_m t}\,\mathrm{d}t$ and satisfies $p_{-n}=p_n^{*}$. The experimentally measured Fourier coefficients $C_n$ therefore determine the relative complex amplitudes of $p_n$, apart from an overall pump-strength calibration.

The evolution of the ensemble spin polarization $\bm{F}=(F_x,F_y,F_z)^T$ is described by the Bloch equation \cite{opticalpumpingreview}
\begin{equation}
	\label{eqn:BlochEquation}
	\dot{\bm{F}}
	=\bm{\omega}_0\times\bm{F}
	-\xi p(t)\bm{F}
	+\kappa p(t)\hat{\bm{s}}
	-\bm{F}_{\perp}/T_2
	-\bm{F}_{\parallel}/T_1,
\end{equation}
where $\bm{\omega}_0=\gamma\bm{B}_0$ and $\hat{\bm{s}}$ denotes the pumping direction. The coefficients $\xi$ and $\kappa$ characterize the pump-induced relaxation and spin-polarization rate, respectively, with the overall scale of $p(t)$ absorbed into these coefficients. The intrinsic transverse and longitudinal relaxation times are denoted by
$T_2$ and $T_1$. For the present geometry, $\bm{B}_0\parallel \hat{e}_z$ and $\hat{\bm{s}}\parallel \hat{e}_x$, so that $\bm{F}_{\parallel}=F_z\hat{e}_z$ and $\bm{F}_{\perp}=F_x\hat{\bm{x}}+F_y\hat{e}_y$.

After introducing the complex transverse polarization $F_{+}=F_x+iF_y$ and $W_0=1/T_2$, Eq.~\ref{eqn:BlochEquation} reduces to
\begin{equation}
	\label{eqn:simplifiedBloch}
	\dot{F}_{+}
	=
	\left[i\omega_0-W_0-\xi p(t)\right]F_{+}
	+\kappa p(t).
\end{equation}

In the periodic steady state, the atomic polarization inherits the modulation period and can therefore be expanded as $F_{+}(t)=\sum_n F_{+,n}e^{in\omega_m t}$. Substituting this expansion together with Eq.~\ref{eqn:FourierExpand} into Eq.~\ref{eqn:simplifiedBloch} yields the coupled Fourier equations:
\begin{equation}
	\label{eqn:BlochinFourier}
	\left[
	W_0+\xi p_0+i(n\omega_m-\omega_0)
	\right]F_{+,n}
	+
	\xi\sum_{q\neq0}p_qF_{+,n-q}
	=
	\kappa p_n .
\end{equation}
To cast Eq.~\ref{eqn:BlochinFourier} into a more compact form and make its resonance structure explicit, we define the harmonic detuning $\delta_n=n\omega_m-\omega_0$ and the complex inverse susceptibility $L_n=W_0+\xi p_0+i\delta_n$, yielding:
\begin{equation}
	\label{eqn:SimplifiedBlochinFourier}
	L_nF_{+,n}
	+
	\xi\sum_{q\neq0}p_qF_{+,n-q}
	=
	\kappa p_n .
\end{equation}

As has been explained in Sec.\ref{sec:principle}, the experimentally detected differential signal is proportional to $F_y$. Therefore its $n$th Fourier coefficient, denoted as$S_n$, satisfies 
\begin{equation}
	\label{eqn:SignalCoefficient}
	S_n
	\propto F_{y,n}
	=
	\frac{
		F_{+,n}-F_{+,-n}^{*}
	}{2i}.
\end{equation}
Near the positive-frequency resonance, where the counter-rotating component $F_{+,-n}^{*}$ is negligible, this relation reduces to $S_n\propto F_{+,n}/(2i)$.

For numerical tractability, we truncate the harmonic basis at order $N$, retaining the Fourier components with $|n|\leq N$ that contain the dominant spectral weight. Eq.~\ref{eqn:SimplifiedBlochinFourier} is thereby reduced to the finite-dimensional Floquet system:
\begin{widetext}
	\begin{equation}
		\label{eqn:MatrixFloquet}
		\left[
		\underbrace{
			\begin{pmatrix}
				L_{-N} & 0 & 0 & \cdots & 0 \\
				0 & L_{-(N-1)} & 0 & \cdots & 0 \\
				0 & 0 & L_{-(N-2)} & \cdots & 0 \\
				\vdots & \vdots & \vdots & \ddots & \vdots \\
				0 & 0 & 0 & \cdots & L_N
			\end{pmatrix}
		}_{\bm{L}}
		+
		\xi
		\underbrace{
			\begin{pmatrix}
				0 & p_{-1} & p_{-2} & \cdots & p_{-2N} \\
				p_1 & 0 & p_{-1} & \cdots & p_{-(2N-1)} \\
				p_2 & p_1 & 0 & \cdots & p_{-(2N-2)} \\
				\vdots & \vdots & \vdots & \ddots & \vdots \\
				p_{2N} & p_{2N-1} & p_{2N-2} & \cdots & 0
			\end{pmatrix}
		}_{\bm{V}}
		\right]
		\underbrace{
			\begin{pmatrix}
				F_{+,-N} \\
				F_{+,-(N-1)} \\
				F_{+,-(N-2)} \\
				\vdots \\
				F_{+,N}
			\end{pmatrix}
		}_{\bm{F}_{+}}
		=
		\kappa
		\underbrace{
			\begin{pmatrix}
				p_{-N} \\
				p_{-(N-1)} \\
				p_{-(N-2)} \\
				\vdots \\
				p_N
			\end{pmatrix}
		}_{\bm{p}} .
	\end{equation}
\end{widetext}
Here, the diagonal matrix $\bm{L}$ describes the uncoupled response of each harmonic, whereas the off-diagonal term $\xi\bm{V}$ accounts for Floquet mixing between different harmonics. The Fourier components of the atomic polarization can then be obtained from
\begin{equation}
	\label{eqn:CompactForm}
	\bm{F}_{+}
	=
	\kappa\left(\bm{L}+\xi\bm{V}\right)^{-1}\bm{p}
	=
	\kappa\bm{M}^{-1}\bm{p},
\end{equation}
where $\bm{M}=\bm{L}+\xi\bm{V}$. When the off-diagonal Floquet coupling is sufficiently weak such that $\rho(\xi\bm{L}^{-1}\bm{V})<1$, where $\rho$ denotes the spectral radius, the inverse matrix can be expanded as a Neumann series,
\begin{align}
	\label{eqn:NeumannSeries}
	\bm{M}^{-1}
	&=
	\sum_{\ell=0}^{\infty}
	(-\xi)^{\ell}
	\left(\bm{L}^{-1}\bm{V}\right)^{\ell}
	\bm{L}^{-1}
	\\
	&=
	\bm{L}^{-1}
	-\xi\bm{L}^{-1}\bm{V}\bm{L}^{-1}
	+\xi^2\bm{L}^{-1}\bm{V}\bm{L}^{-1}
	\bm{V}\bm{L}^{-1}
	+\mathcal{O}(\xi^3). \nonumber
\end{align}
Accordingly, the solution can be decomposed as $\bm{F}_{+}=\sum\nolimits_{\ell=0}^{\infty}\bm{F}_{+}^{(\ell)}$, where
\begin{equation}
	\bm{F}_{+}^{(\ell)}
	=
	(-\xi)^{\ell}\kappa
	\left(\bm{L}^{-1}\bm{V}\right)^{\ell}
	\bm{L}^{-1}\bm{p},
\end{equation}
In the \emph{independent-harmonic response} limit, where the pump light is sufficiently weak that the off-diagonal coupling $\xi\bm{V}$ can be neglected, the response reduces to $\bm{F}_{+}\simeq\bm{F}_{+}^{(0)}=\kappa\bm{L}^{-1}\bm{p}$, or, equivalently, $F_{+,n}^{(0)}=\kappa p_n/L_n$. Combining this result with Eq.~\ref{eqn:SignalCoefficient} and applying the rotating-wave approximation near the positive-frequency resonance gives
\begin{equation}
	\label{eqn:zerothordersolution}
	\mathcal{H}_n
	\equiv
	\frac{S_n}{p_n}
	\propto
	\frac{\kappa/(2i)}
	{W_0+\xi p_0+i(n\omega_m-\omega_0)}.
\end{equation}
The above expression indicates that the normalized atomic response to the $n$th harmonic drive exhibits a Lorentzian line shape with respect to $\delta_n=n\omega_m-\omega_0$, with its linewidth determined by the effective transverse relaxation rate $\Gamma=W_0+\xi p_0$. Thus, in the \emph{independent-harmonic response} regime, the pump waveform determines only the complex driving amplitudes $p_n$, while the normalized response $\mathcal{H}_n$ is independent of the detailed harmonic structure and responses from different harmonic orders are therefore expected to collapse onto the same complex Lorentzian.

As the pump strength increases, non-negligible \emph{Floquet mixing} causes
deviations from the universal Lorentzian response Eq.~\ref{eqn:zerothordersolution}. To elucidate this mechanism, we consider the leading correction in the Neumann expansion, i.e. retaining only the first order off-diagonal coupling term $\xi\bm{V}$. For a Floquet basis truncated at $|n|\leq N$, this correction is given by
\begin{equation}
	\label{eqn:FirstOrderFloquetCorrection}
	F_{+,n}^{(1)}
	=
	-\frac{\kappa\xi}{L_n}
	\sum_{\substack{q\ne0\\|n-q|\leq N}}
	\frac{p_qp_{n-q}}{L_{n-q}}.
\end{equation}

As an illustrative example, retaining pump harmonics up to the
second order ($N=2$) yields
\begin{equation}
	\label{eqn:SecondHarmonicResponse}
	F_{+,2}
	\simeq
	\frac{\kappa p_2}{L_2}
	-
	\frac{\kappa\xi}{L_2}
	\left(
	\frac{p_1^2}{L_1}
	+
	\frac{p_2p_0}{L_0}
	\right).
\end{equation}
Remarkably, even when $p_2=0$, a finite second-harmonic response can arise from mixing of the fundamental harmonic, with $F_{+,2}\simeq-\kappa\xi p_1^2/(L_2L_1)$. Generally speaking, the mixing terms introduce contributions from other harmonics. Rearranging the Floquet equation Eq.~\ref{eqn:SimplifiedBlochinFourier}, within the near-resonant approximation, we get:
\begin{equation}
	\label{eqn:FloquetCorrection}
	\mathcal H_n \propto
	\frac{\kappa}{2iL_n}(1-\mathcal B_n),
	\qquad
	\mathcal B_n =
	\frac{\xi}{\kappa p_n}
	\sum_{q\neq0}p_qF_{+,n-q},
\end{equation}
for $p_n\neq0$, where $F_{+,n-q}$ denotes the full Floquet solution. The complex correction $\mathcal B_n$ introduces harmonic-dependent modifications to both the amplitude and phase of $\mathcal H_n$. Coupling to other harmonics can also modify the response away from the $n$th-harmonic resonance, producing \emph{non-Lorentzian features} and apparent \emph{off-resonant backgrounds}. Consequently, the normalized responses no longer necessarily collapse onto a common Lorentzian profile. The following section provides experimental observations that qualitatively support the effects described by Eq.~\ref{eqn:FloquetCorrection}.

\section{\label{sec:sixwaveform}From independent harmonic response to Floquet mixing}
\begin{figure}[htb!]
	\centering
	\includegraphics[width=\linewidth]{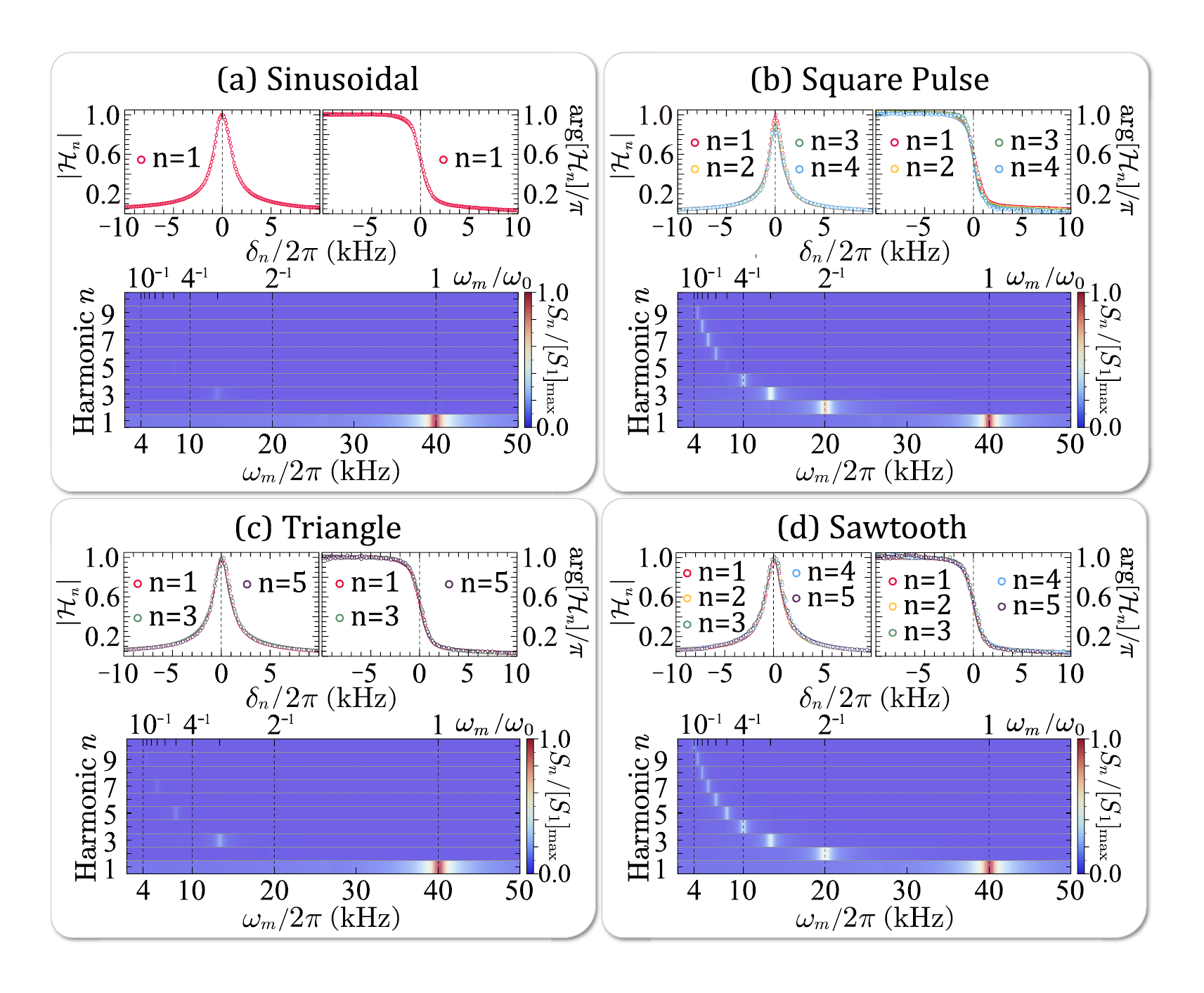}
	\caption{\textbf{Harmonic-resolved atomic responses under weak-pump (independent harmonic response) condition} for (a) sinusoidal, (b) square-pulse ($D=0.2$), (c) triangular, and (d) rising-sawtooth modulation. The upper panels of each waveform show the amplitudes and phases of the normalized responses as functions of $\delta_n$, where open circles represent experimental data and solid curves denote the corresponding fits using Eq.~\ref{eqn:fitmodel}. The density plots show the measured Fourier amplitudes $|S_n|/\max\left(|S_1|\right)$ versus modulation frequency.
	}
	\label{Fig:weakpump}
\end{figure}
To investigate interharmonic Floquet mixing, we compared the responses to six pump waveforms under weak- and strong-pump conditions. For weak pumping, the vapor temperature was set to $130,^\circ\mathrm{C}$, with an independently measured intrinsic relaxation rate of $W_0=1/T_2\approx1.3~\mathrm{kHz}$, whose relatively large value helps suppress harmonic mixing. The pump-induced relaxation rate was measured separately at the peak intensity of each waveform and scaled by its mean-to-peak intensity ratio to determine $\xi p_0$, yielding the reference linewidth $\sigma_{\rm ind}=W_0+\xi p_0$.

The weak-pump responses for different waveforms are presented in Fig.~\ref{Fig:weakpump}. The lower density plots show the amplitudes of individual response harmonics as functions of $f_m$, with the observed harmonic orders reflecting the Fourier spectra of the corresponding waveforms (see Fig.~\ref{Fig:differentwaveform}). Each directly driven harmonic exhibits a resonance at $n\omega_m=\omega_0$, and such higher-order resonances may also be exploited in optical magnetometry \cite{Gerginov2017}. The upper plots display the amplitudes and phases of the pump-normalized responses, which approximately collapse onto common resonance profiles across different harmonic orders, consistent with the independent-harmonic approximation [Eq.~\ref{eqn:zerothordersolution}]. Notably, square-pulse modulation with $D=0.2$ produces narrower resonances than the other waveforms, consistent with its lower mean pump-induced relaxation rate $\xi p_0$ at comparable peak pump intensities.

For quantitative analysis, we separately fit the amplitude spectrum $|\mathcal H_n|$ of each directly driven harmonic using the following Lorentzian-based function:
\begin{equation}
	\label{eqn:fitmodel}
	f_n(\Delta_n)=K_n
	\left[
	\frac{A_{{\rm fit},n}}
	{\sqrt{1+(\Delta_n/\sigma_{{\rm fit},n})^2}}
	+(1-A_{{\rm fit},n})
	\right],
\end{equation}
Here,  $\Delta_n=\delta_n/(2\pi)$, $K_n$ denotes the fitted response amplitude at resonance, $A_{{\rm fit},n}$ specifies the fractional resonant contribution, and $\sigma_{{\rm fit},n}$ characterizes the resonance width. The remaining contribution, $K_n(1-A_{{\rm fit},n})$, represents a phenomenological amplitude background. Within the independent-harmonic regime, harmonics driven by the same pump waveform
are expected to share a common normalized resonance amplitude. We therefore quantify their inter-harmonic dispersion through
\begin{equation}
	\label{eqn:sK}
	s_K=
	\frac{1}{\overline K}
	\sqrt{
		\frac{1}{|\mathcal N|-1}
		\sum_{n\in\mathcal N}
		(K_n-\overline K)^2
	},
	\qquad
	\overline K=
	\frac{1}{|\mathcal N|}
	\sum_{n\in\mathcal N}K_n,
\end{equation}
where $\mathcal N$ contains the allowed pump harmonic orders up to the fifth order. The quantity $s_K$ thus characterizes the relative dispersion of the pump-normalized resonance amplitudes.
\begin{table}[thb!]
	\caption{Spectral fitting results under weak-pump condition. $A_{\rm fit}$ and $\sigma_{\rm fit}$ denote the means of different harmonics, with standard deviations given where significant. Uncertainties below the adopted precision are omitted. Here, $\sigma_{\rm ind}$ is predicted from independently measured relaxation rates, while $s_K$ quantifies the relative dispersion of $K_n$ among the harmonics under consideration ($n\leq 5$).}
	\label{tab:weakpump}
	\begin{ruledtabular}
		\begin{tabular}{lllll}
			\multicolumn{1}{l}{Waveform}\footnote{Pulse ($D$) denotes a square pulse with duty cycle $D$. ``Ris. Saw." and ``Fall. Saw." denote rising and falling sawtooth waveforms, respectively.}&
			\multicolumn{1}{l}{$A_{\rm{fit}}$}&
			\multicolumn{1}{l}{$\sigma_{\rm{fit}}\, (\mathrm{kHz})$}&
			\multicolumn{1}{l}{$\sigma_{\rm{ind}}\, (\mathrm{kHz})$}&
			\multicolumn{1}{l}{$s_{\rm{K}}\, (\% )$}\\
			\hline
			Sinusoidal\footnote{Only the fundamental harmonic is considered for sinusoidal modulation.} & 1.00 & 4.2 & 3.3$\pm$0.5 & --\\
			Pulse (0.5) & 1.00 & 4.5$\pm$0.2 & 3.9$\pm$0.5 & $\sim10^{-5}$ \\
			Pulse (0.2) & 1.00 & 3.1$\pm$0.2 & 2.3$\pm$0.4 & 4.9\\
			Triangle & 1.00 & 4.2$\pm$0.2 & 3.2$\pm$0.4 & 0.1\\
			Ris. Saw. & 1.00 & 4.2$\pm$0.2 & 3.0$\pm$0.5 &  $\sim10^{-5}$ \\
			Fall. Saw. & 1.00 & 4.6$\pm$0.2 & 4.1$\pm$0.5 & $\sim10^{-5}$ \\
		\end{tabular}
	\end{ruledtabular}
\end{table}

The fitting results for weak pump condition are summarized in Tab.~\ref{tab:weakpump}. Here, near-unity values of $A_{\rm fit}$ and generally small $s_K$ consistent with the approximately collapsed normalized responses. Notably, the largest dispersion, $s_K=4.9\%$, occurs for pulse modulation with $D=0.2$, reflecting an increased sensitivity to spectral imperfections. Moreover, the fitted characteristic widths follow the waveform-dependent trends of the independently determined linewidths. Nevertheless, $\sigma_{\rm fit}$ exceeds $\sigma_{\rm ind}$ for all six waveforms, indicating a residual broadening not captured by the current simplified model. This discrepancy may arise from temporal fluctuations in experimental conditions.
\begin{figure}[htb!]
	\centering
	\includegraphics[width=\linewidth]{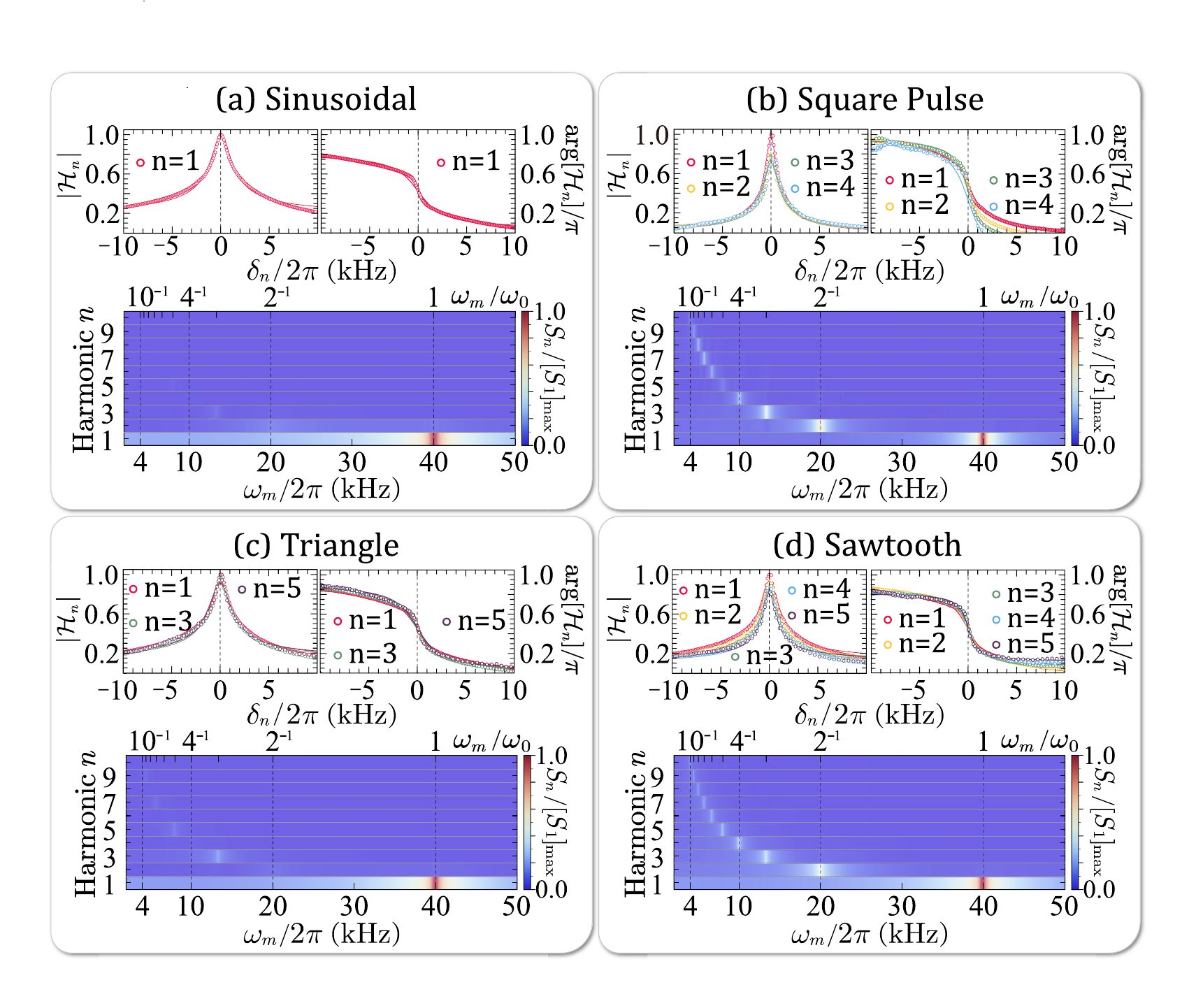}
	\caption{\textbf{Harmonic-resolved atomic responses under strong-pump (Floquet Mixing) condition} for (a) sinusoidal, (b) square-pulse ($D=0.2$), (c) symmetric triangular, and (d) falling-sawtooth modulation. The plotting and normalization conventions follow Fig.~\ref{Fig:weakpump}.
	}
	\label{Fig:strongpump}
\end{figure}

For the strong-pump measurements, the vapor temperature was increased to $162\,^\circ\mathrm{C}$, corresponding to a reduced $W_0\approx0.7~\mathrm{kHz}$ \cite{SpinExchangeNarrowing}. This weaker intrinsic relaxation, together with stronger optical pumping, enhances the relative importance of inter-harmonic coupling. Under this condition, the harmonic responses exhibit more pronounced deviations from the uncoupled description, as shown in Fig.~\ref{Fig:strongpump}. Although the Fourier-amplitude density plots retain the principal features associated with the pump spectra, the pump-normalized responses no longer collapse onto common profiles. In particular, enhanced off-resonant amplitudes and harmonic-dependent variations in both amplitude and phase become apparent, consistent with the increased inter-harmonic coupling predicted by the Fourier--Floquet model.

These deviations are further quantified in Tab.~\ref{tab:strongpump}. Under current condition, $A_{\rm fit}$ decreases to $0.81$--$0.98$, indicating appreciable background contributions. Meanwhile, the inter-harmonic dispersion $s_K$ generally increases, reaching $15.8\%$ for pulse modulation with $D=0.2$, compared with $4.9\%$ under weak pumping. The different values of $s_K$ across waveforms further reveal the waveform-dependence deviations. More strikingly, $\sigma_{\rm fit}$ falls below $\sigma_{\rm ind}$ for all six waveforms, in contrast to the trend observed under weak pumping. Since $\sigma_{\rm ind}$ is independently determined from $W_0+\xi p_0$, this reversal cannot be explained by the mean pump-induced relaxation. The narrower-than-reference linewidths may be associated with the spatially inhomogeneous optical pumping and the resulting averaging of atomic responses. This effect is reminiscent of diffusion-induced Ramsey narrowing in its spectral manifestation, although the underlying physical mechanisms differ ~\cite{RamseyNarrowing06}. The role of spatial averaging is examined in the following section.

\begin{table}[htb!]
	\caption{
		Spectral fitting results under strong-pump condition. The fitting procedure and statistical conventions follow Tab.~\ref{tab:weakpump}.
	}
	\label{tab:strongpump}
	\begin{ruledtabular}
		\begin{tabular}{lllrr}
			\multicolumn{1}{l}{Waveform}&
			\multicolumn{1}{l}{$A_{\rm{fit}}$}&
			\multicolumn{1}{l}{$\sigma_{\rm{fit}}\, (\mathrm{kHz})$}&
			\multicolumn{1}{l}{$\sigma_{\rm{ind}}\, (\mathrm{kHz})$}&
			\multicolumn{1}{l}{$s_{\rm{K}}\, (\% )$} \\
			\hline
			Sinusoidal & 0.81 & 7.2 & 10.3$\pm$0.6 & - \\
			Pulse (0.5) & 0.91$\pm$0.03 & 7.1$\pm$1.3 & 10.4$\pm$0.5 & 3.9\\
			Pulse (0.2) & 0.98$\pm$0.01 & 4.3$\pm$0.8 & 5.3$\pm$0.4 & 15.8\\
			Triangle & 0.85$\pm$0.02 & 6.6$\pm$0.2 & 10.8$\pm$0.6 & 1.2\\
			Ris. Saw. & 0.84$\pm$0.02 & 6.6$\pm$1.4 & 9.7$\pm$0.7 & 7.2\\
			Fall. Saw. & 0.87$\pm$0.02 & 5.0$\pm$0.7 & 9.5$\pm$0.7 & 5.7\\
		\end{tabular}
	\end{ruledtabular}
\end{table}

\section{\label{sec:spatialparameter}Phenomenological Parameter for Spatial Geometry}
\begin{figure}[htb!]
	\centering
	\includegraphics[width=\linewidth]{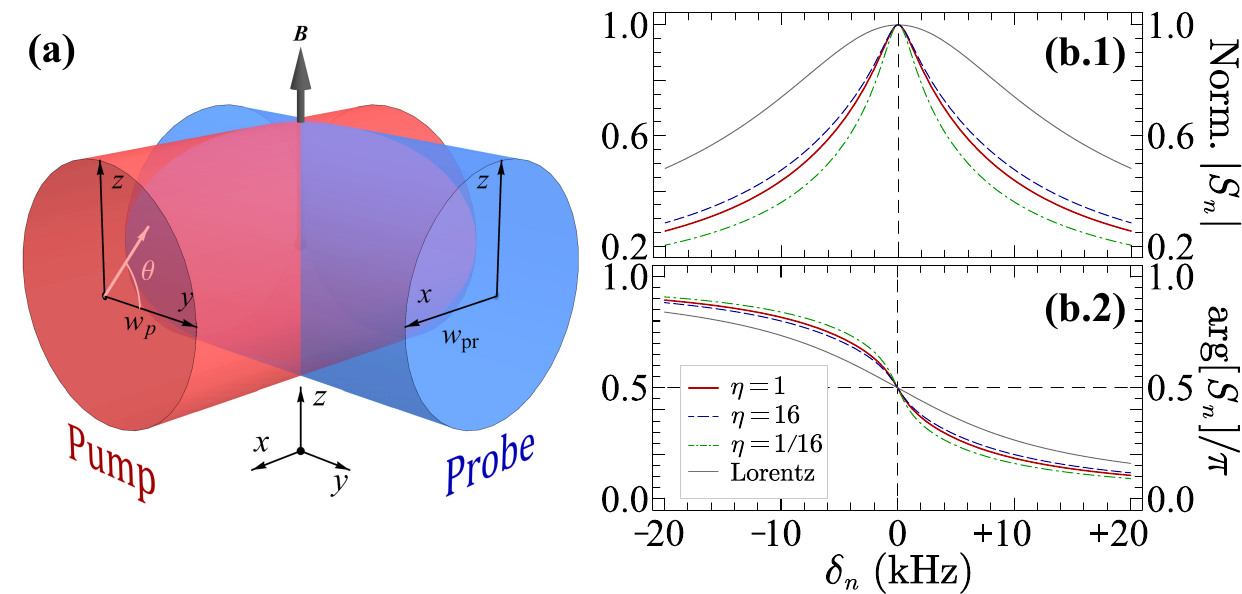}
	\caption{\textbf{Spatial averaging of the atomic harmonic response.} (a) Schematic of the orthogonal pump--probe geometry in the adopted Cartesian coordinate. (b) Simulated normalized amplitude (b.1) and phase (b.2) of the spatially averaged response $S_n$ as functions of harmonic detuning for different $\eta$. The gray curves represent the raw Lorentzian response without spatial averaging. Simulation parameters: \(W_0 = 1~\mathrm{kHz}\) and \(p_0 =10~\mathrm{kHz}\).
	}
	\label{Fig:distribution}
\end{figure}
To take the spatial distribution of optical pumping and detecting into consideration, the transverse profiles of the pump and probe beams are described by Gaussian functions. As illustrated in Fig.~\ref{Fig:distribution} (a), their respective $1/e^2$ radii, $w_p$ and $w_{pr}$, define the spatial distributions:
\begin{equation}
	\label{eqn:gaussiandistribution}
	\begin{aligned}
		\text{pump:}&\quad u_p(y,z) =
		\exp\left[-2(y^2+z^2)/ w_p^2\right],\\
		\text{probe:}&\quad u_{pr}(x,z) =
		\exp\left[-2(x^2+z^2)/w_{pr}^2\right].
	\end{aligned}
\end{equation}
by introducing the dimensionless radial coordinate
\begin{equation}
	\label{eqn:dimensionlesst}
	t\equiv 2(y^2+z^2)/w_p^2,
\end{equation}
the spatial dependence of the optical pumping can then be incorporated into Eq.~\ref{eqn:CompactForm} through the substitution:
\begin{equation}
	\label{eqn:includepumpweight}
	\bm p\rightarrow u_p\bm p=e^{-t}\bm p.
\end{equation}
Meanwhile, by defining the effective geometric parameter:
\begin{equation}
	\label{eqn:eta}
	\eta=w_p^2/(2w_{pr}^2),
\end{equation}
the probe weighting along the common transverse coordinate $z$ can similarly be expressed as
\begin{equation}
	\label{eqn:probeweighting}
	\exp\left(-2z^2/w_{pr}^2\right)
	=\exp\left(-2\eta t\sin^2\theta\right).
\end{equation}
Here, $(r,\theta)$ are polar coordinates in the $yz$ plane, with $r=\sqrt{y^2+z^2}$ and $z=r\sin\theta$. Integrating Eq.~\ref{eqn:probeweighting} over the whole plane then yields:
\begin{equation}
	\label{eqn:integratepumpplane}
	\int_0^{2\pi}
	e^{-2\eta t\sin^2\theta}\,\mathrm d\theta
	=2\pi e^{-\eta t}I_0(\eta t),
\end{equation}
where $I_0$ is the modified Bessel function of the first
kind of order zero.

Combining Eqs.~\ref{eqn:includepumpweight} and \ref{eqn:integratepumpplane} then reduces the spatial average to a one-dimensional integral. We first evaluate the spatially averaged line shape in the independent-harmonic limit. Applying the spatial weights to Eq.~\ref{eqn:zerothordersolution} gives
 \begin{equation}
 	\label{eqn:zerothordersolutionwithlightcorrection}
 	\mathcal{H}_{n}\propto \frac{\kappa}{2i}\int_{0}^{\infty}{\frac{e^{-(1+\eta)t}I_0(\eta t)}{W_0+\xi p_0 e^{-t}+i\left(n\omega_m - \omega_0\right)}\mathrm{d}t}.
 \end{equation}
In the weak-pumping limit, $\xi p_0\rightarrow0$, Eq.~\ref{eqn:zerothordersolutionwithlightcorrection} reduces to the standard complex Lorentzian, apart from an overall scaling factor of $1/\sqrt{2\eta+1}$. Therefore, spatial averaging leaves the line shape essentially unchanged in this regime, consistent with the nearly Lorentzian responses observed in Fig.~\ref{Fig:weakpump}. On the other hand, the amplitude and phase profiles calculated from Eq.~\ref{eqn:zerothordersolutionwithlightcorrection} under strong pump condition are depicted in Fig.~\ref{Fig:distribution}(b). We find that as $\eta$ decreases, the response narrows near the resonance, with a correspondingly steeper slope of phase. By contrast, the complex Lorentzian is recovered when the pump intensity is effectively uniform. This simulated narrowing suggests that spatial averaging may help explain why $\sigma_{\rm fit}$ falls below the power-broadened reference width $\sigma_{\rm ind}$ under strong pumping (Table~\ref{tab:strongpump}). However, it cannot account for either the off-resonant background or the loss of harmonic-response collapse under strong pumping. Accordingly, we extend the simplified model Eq.~\ref{eqn:zerothordersolutionwithlightcorrection} to include inter-harmonic Floquet coupling, yielding
\begin{equation}
	\label{eqn:SpatialFloquetResponse}
	S_n \propto -i\pi\int_{0}^{\infty}{e^{-\eta t}I_0(\eta t)\left(F_{+,n}-F_{+,-n}^{*}\right)\rm{d} t}
\end{equation}
where the local coefficients $F_{+,\pm n}(t)$ are obtained from Eq.~\eqref{eqn:CompactForm} using the pump weighting in Eq.~\eqref{eqn:includepumpweight}.

In a realistic vapor cell, however, the detected response is influenced by spatial inhomogeneities beyond the idealized geometry, including the cell-boundary effects, imperfect pump--probe overlap and residual magnetic-field gradients. Rather than modeling these contributions individually, we regard $\eta$ as an effective phenomenological spatial parameter that incorporates
their net influence on the ensemble-averaged response. In the following section, this spatially averaged Floquet model is tested using a tailored pump waveform with a deliberately sparse Fourier spectrum.

\section{\label{sec:engineeredwaveform}Interplay between Floquet Mixing and Spatial Averaging}
\begin{figure*}[th!]
	\centering
	\includegraphics[width=\linewidth]{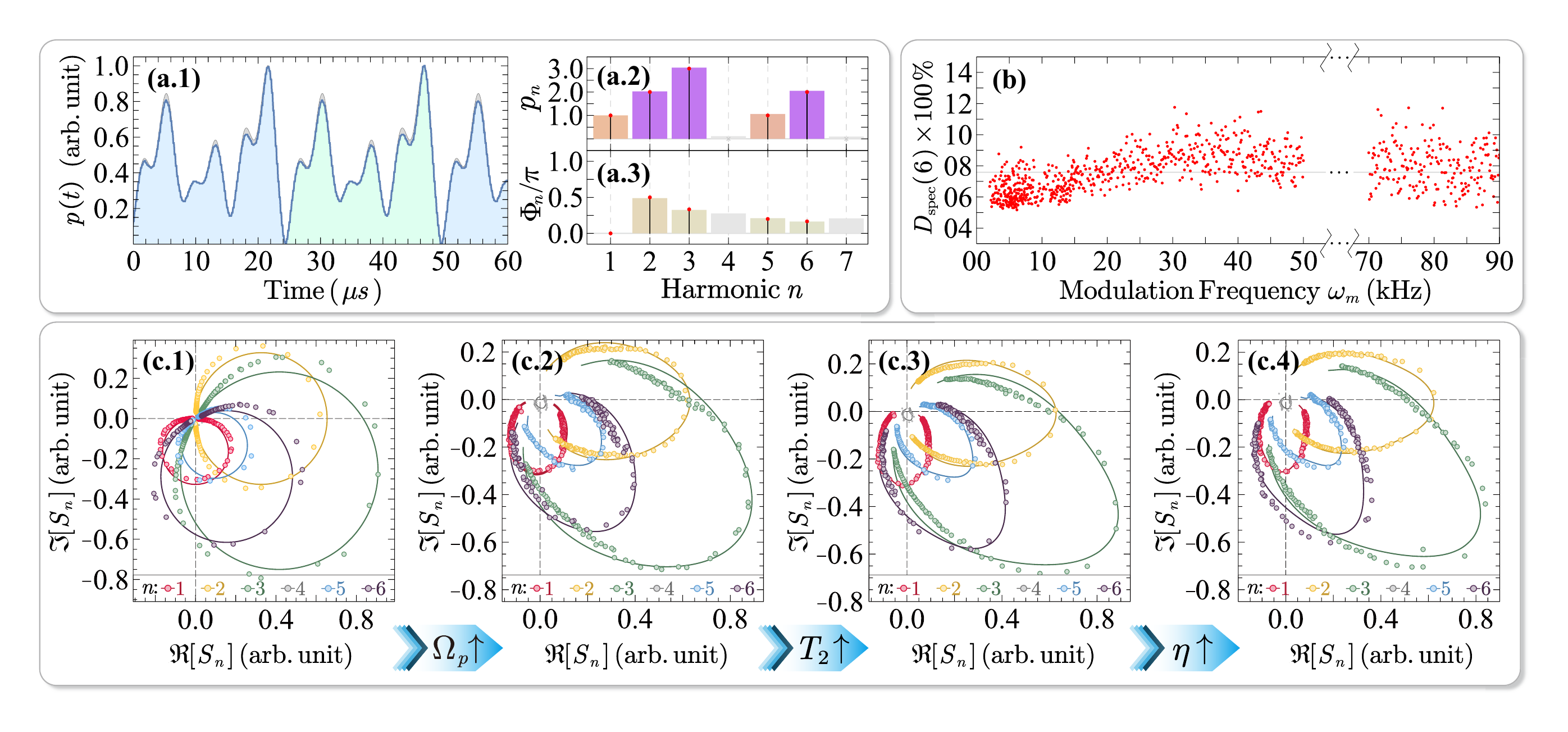}
	\caption{\textbf{Characterization of the tailored pump waveform and the corresponding atomic Floquet response.} (a.1) Ideal waveform at 100\% modulation depth (light gray) and the realistic waveform registered by the fast photodiode (highlighted in blue and green). (a.2,a.3) Fourier amplitudes $p_n$ and relative phases $\Phi_n/\pi$, respectively, at $f_m=f_0=40~\mathrm{kHz}$. Bars represent the measured Fourier components and red markers denote the target values. (b) Spectral deviation $D_{\rm spec}(6)$ as a function of modulation frequency. (c.1--c.4) Complex harmonic responses $\mathcal H_n$ for $n=1,\ldots,6$ under different experimental conditions. Open circles show the experimental data. The solid curves inc(c.1)--(c.4) are numerical results from the spatially averaged Floquet model of Eq.~\eqref{eqn:SpatialFloquetResponse}, with $W_0+\xi p_0$ independently determined. 
	}
	\label{Fig:engineeredwaveform}
\end{figure*}
To further test the proposed model, we adopt a tailored pump waveform with deliberately designed Fourier spectrum:
\begin{equation}
	\label{eqn:targetwaveform}
	\begin{aligned}
		p(\tau)=\frac{1}{14}\Bigg\{
		&\overbrace{\cos(2\pi\tau)}^{n=1}
		+\overbrace{2\cos\left(4\pi\tau+\frac{\pi}{2}\right)}^{n=2}
		+\overbrace{3\cos\left(6\pi\tau+\frac{\pi}{3}\right)}^{n=3}
		\\
		&+\underbrace{\cos\left(10\pi\tau+\frac{\pi}{5}\right)}_{n=5}
		+\underbrace{2\cos\left(12\pi\tau+\frac{\pi}{6}\right)}_{n=6}
		\Bigg\}+C_0 .
	\end{aligned}
\end{equation}
where $\tau=f_m t_{\rm lab}$ is the normalized time and $C_0=0.445$ corresponds to approximately 100\% modulation depth. This waveform contains only a finite set of harmonics, allowing the Floquet space to be truncated to a finite number of coupled orders. In addition, the fourth harmonic is suppressed by design, so that any response emerging at this order can serve as a direct signature of Floquet mixing. Finally, the fundamental component is not dominant and the retained harmonics have prescribed relative phases, providing a stringent test of the model's validity and generality through the harmonic-resolved complex response.

The realistic modulation waveform is first compared with the ideal one. As shown in Fig.~\ref{Fig:engineeredwaveform}(a.1), the measured pump waveform closely follows the target profile. At $f_m=f_0=40~\mathrm{kHz}$, the Fourier amplitudes and relative phases (Figs.~\ref{Fig:engineeredwaveform}(a.2) and (a.3)) of registered waveform reproduce the prescribed components at $n=1,2,3,5,$ and $6$, while the fourth harmonic remains suppressed. What's more, across a broad modulation-frequency range, $D_{\rm spec}(6)$ remains small [Fig.~\ref{Fig:engineeredwaveform}(b)], demonstrating the robustness of our waveform-generation method. In the analysis below, the measured complex Fourier coefficients are used as the parameters for theoretical calculations.

Fig.~\ref{Fig:engineeredwaveform}(c) compares the measured harmonic responses (open circles) with the simulation results using Eq.~\ref{eqn:SpatialFloquetResponse} (solid curves) in the complex plane over a detuning range of $-12\leq\delta_n/2\pi\leq12~\mathrm{kHz}$. This representation displays both quadratures simultaneously, highlights near-resonant discrepancies, and reveals departures from the circular trajectory of an ideal complex Lorentzian. Consistent with Tabs.~\ref{tab:weakpump} and \ref{tab:strongpump}, the weak-pump response (Fig.~\ref{Fig:engineeredwaveform}(c.1)) is broader than the prediction from Lorentzian model($\sigma_{\rm fit}>\sigma_{\rm ind}$), whereas the strong-pump responses (Figs.~\ref{Fig:engineeredwaveform}(c.2)--(c.4)) are narrower. The spatially averaged model captures the observed responses and their evolution across different experimental conditions qualitatively.

To quantitatively assess the agreement between experiment and the proposed model, we evaluate the pointwise discrepancy by
\begin{equation}
	\label{eqn:pointerr}
	\epsilon_{n,i}(R)=
	\frac{
		\left|P_{n,i}^{\rm exp}-R P_{n,i}^{\rm theo}\right|
	}{
		\left|P_{n,i}^{\rm exp}\right|
		+R\left|P_{n,i}^{\rm theo}\right|
	} .
\end{equation}
Here, $P_{n,i}^{\rm exp}$ and $P_{n,i}^{\rm theo}$ denote the experimental and calculated complex responses, respectively. The positive real scale factor $R$ is common to all harmonic orders and accounts only for the overall amplitude difference between experiment and calculation. The discrepancies for all the measured points are then combined into the harmonic-resolved error and averaged over the selected orders:
\begin{equation}
	\label{eqn:relativeerr}
	\mathcal E_n(R)=
	\sqrt{
		\frac{1}{\mathcal M_n}
		\sum_{i=1}^{\mathcal M_n}
		\epsilon_{n,i}^{2}(R)
	},
	\quad
	\overline{\mathcal E}(R)=
	\frac{1}{|\mathcal N|}
	\sum_{n\in\mathcal N}\mathcal E_n(R),
\end{equation}
Here, $\mathcal M_n$ is the number of sampled points for the $n$th harmonic and $\mathcal N=\{1,2,3,5,6\}$. The single global scale factor $R$ is determined by minimizing $\overline{\mathcal E}(R)$ over $R>0$.

To assess both the predictive accuracy and the corresponding regime of validity of different descriptions, we compare three models under several experimental conditions: the bare Lorentzian (L), the Floquet mixing without spatial averaging (F), and the spatially averaged Floquet model (S). For each experimental condition, the relaxation parameters $W_0$ and $\xi p_0$ are determined independently, while the measured pump Fourier amplitudes and phases are used as inputs to the calculations. In the spatial model, $\eta$ is determined from the beam sizes. The corresponding discrepancies are denoted by $\overline{\mathcal E}_{\rm L}$, $\overline{\mathcal E}_{\rm F}$, and $\overline{\mathcal E}_{\rm S}$, respectively. The results are summarized in Tab.~\ref{tab:modelcomparison}.
\begin{table}[htb!]
	\caption{
		Comparison of the model discrepancies for the tailored waveform under different experimental conditions. The $\pm$ values represent standard deviations over included harmonics.
	}
	\label{tab:modelcomparison}
	\begin{ruledtabular}
		\begin{tabular}{lrrrrrr}
			\multicolumn{1}{l}{No.}&
			\multicolumn{1}{l}{$W_0\, (\mathrm{kHz})$}&
			\multicolumn{1}{l}{$\xi p_0 \, (\mathrm{kHz})$}&
			\multicolumn{1}{c}{$\eta$}&
			\multicolumn{1}{l}{$\overline{\mathcal{E}}_{\rm{L}}\, (\% )$}&
			\multicolumn{1}{l}{$\overline{\mathcal{E}}_{\rm{F}}\, (\% )$}&
			\multicolumn{1}{l}{$\overline{\mathcal{E}}_{\rm{S}}\, (\% )$} \\
			\hline
			1 & 1.2$\pm$0.3 & 1.2$\pm$0.2 & 2.0 & 8.5$\pm$1.3 & 7.4$\pm$1.2 & 8.2$\pm$1.2\\
			2 & 0.8$\pm$0.1 & 0.6$\pm$0.1 & 2.0 & 7.6$\pm$1.5 & 7.6$\pm$1.5 & 7.7$\pm$1.3\\
			3 & 1.2$\pm$0.2 & 12.8$\pm$0.8 & 2.0 & 13.3$\pm$1.7 & 8.9$\pm$0.6 & 5.1$\pm$1.2\\
			4 & 0.7$\pm$0.1 & 12.9$\pm$0.8 & 2.0 & 13.4$\pm$1.2 & 11.9$\pm$1.2 & 5.2$\pm$0.5\\
			5 & 0.7$\pm$0.1 & 14.6$\pm$0.8 & 16.0 & 11.7$\pm$0.8 & 12.8$\pm$0.9 & 3.7$\pm$0.9\\
		\end{tabular}
	\end{ruledtabular}
\end{table}

At the lower pump strengths of Nos.~1 and 2 ($\xi p_0\sim W_0$), all three models yield comparable discrepancies, with no systematic improvement from spatial averaging. In No.~2, $\overline{\mathcal E}{\rm S}$ is even marginally larger than $\overline{\mathcal E}{\rm L}$. This behavior is consistent with the approach to a complex Lorentzian in the weak-pump limit (Sec.~\ref{sec:sixwaveform}), although spatial averaging alone does not explain the excess linewidth broadening ($\sigma_{\rm fit}>\sigma_{\rm ind}$) observed in the weak-pump measurements (Table~\ref{tab:weakpump}). A marked contrast emerges under strong pumping (Nos.~3--5). The spatially averaged Floquet model yields the smallest mean discrepancies ($\overline{\mathcal E}_{\rm S}=3.7$--$5.2\%$), compared with $8.9$--$12.8\%$ for the spatially uniform models, representing a reduction of approximately $43$--$68\%$. Although Floquet mixing alone improves agreement with experiment to varying degrees, spatial averaging provides a further, consistent reduction in discrepancy, highlighting the importance of both effects in the strong-pump regime.

Residual discrepancies of several percent under strong pumping suggest that the nominal Gaussian geometry may not fully capture the experimental spatial weighting. We therefore introduce an effective geometric parameter $\eta_{\rm eff}$, jointly optimized with the overall scale factor $R$. Relative to the nominal $\eta$, smaller (larger) values of $\eta_{\rm eff}$ favor contributions from weakly pumped (strongly pumped central) regions. This phenomenological adjustment may account for unmodeled effects such as imperfect beam overlap, atomic transport, or finite-aperture clipping.

\section{\label{sec:conclusion}Conclusion}
In conclusion, we have extended and examined a Fourier--Floquet framework for atomic responses to arbitrary periodic optical pumping. By incorporating the measured pump spectrum, independently determined relaxation rates, and spatial averaging, the model unifies the independent-harmonic and Floquet-mixing regimes within a common description. Its applicability was further tested using a tailored waveform with prescribed Fourier amplitudes and phases, providing a controlled probe of spectral composition, interharmonic coupling, and spatial effects.

The frequency-domain formulation also provides a natural route toward more general excitation schemes. Periodic responses can be reconstructed from their complex Fourier components, while direct temporal propagation offers an extension to transient and nonperiodic pumping \cite{Gerginov2017}. Driving with multiple independent frequencies presents a further direction for investigating coupled atomic dynamics \cite{Floquet2021PRA,Geng26PRA}. A complementary experimental test would compare Gaussian and flat-top pump profiles \cite{Yuan2024FlatTop,Xu2025Homogenization}: for ideally uniform pumping throughout the probed volume, spatial averaging reduces to an overall amplitude factor, while Floquet mixing persists. Future work may also incorporate atomic transport and finite-cell effects explicitly and explore waveform optimization for harmonic-resolved sensing of magnetic, radio-frequency, and microwave fields.

\section*{Data Availability}
The data that support the findings of this study are available from the corresponding author upon reasonable request.
\begin{acknowledgements}
We acknowledge the fundings from National Natural Science Foundation of China (Grant Nos. 12304561, 12304391, 12447128 and 92576209), Anhui Province Science and Technology Innovation Overcoming Plan Project (Grant No. 202423s06050002) and HeFei University of Technology (Grant No. JZ2026HGTS0249).
\end{acknowledgements}

\bibliography{periodicwaveformengineering}

\end{document}